\documentclass[aps,prd,twocolumn,showpacs,preprintnumbers,amsmath,amssymb,showpacs,showkeys]{revtex4}
\usepackage{graphicx}
\usepackage{dcolumn}
\usepackage{bm}
\usepackage{amssymb}
\usepackage{indentfirst}
\usepackage{epsfig}
\usepackage{epsf}
\usepackage{graphicx}

\newcommand{\eq}{\begin{eqnarray}}
\newcommand{\en}{\end{eqnarray}}
\newcommand{\bfb}{{\bf b}_{\perp}}
\newcommand{\bfk}{{\bf k}_{\perp}}

\newcommand{\ra}{\rangle}
\newcommand{\la}{\langle}

\begin{document}

\title{Generalized parton distributions in an AdS/QCD hard-wall model}

\author{Alfredo Vega$^{1}$,
        Ivan Schmidt$^{1}$,
        Thomas Gutsche$^{2}$ and
        Valery E. Lyubovitskij$^{2}$\footnote{On leave of absence
from Department of Physics, Tomsk State University, 634050 Tomsk, Russia}
\vspace*{1.2\baselineskip}}

\affiliation{$^{1}$Departamento de F\'\i sica y Centro Cient\'ifico y 
Tecnol\'ogico de Valpara\'iso,\\  
Universidad T\'ecnica Federico Santa Mar\'\i a,\\
Casilla 110-V, Valpara\'\i so, Chile \\
\vspace*{.2\baselineskip} \\
$^{2}$ Institut f\"ur Theoretische Physik,
Universit\"at T\"ubingen,\\
Kepler Center for Astro and Particle Physics, \\
Auf der Morgenstelle 14, D--72076 T\"ubingen, Germany
\vspace*{.8\baselineskip}}

\date{\today}

\begin{abstract}

We use a matching procedure of sum rules relating the electromagnetic 
form factors to generalized parton distributions (GPDs) and AdS modes.
In this way, in the framework of an AdS/QCD hard-wall model,
the helicity-independent GPDs of quarks for the nucleon in 
the zero skewness case are calculated. 

\end{abstract}



\preprint{USM-TH-297}

\maketitle

\section{Introduction}

Light Front Holography (LFH)~\cite{Brodsky:2006uqa,Brodsky:2008pg,%
Brodsky:2003px} is a semiclassical approximation to QCD based on the 
gauge/gravity duality that provides a precise mapping of the string 
modes $\Phi(z)$ in the anti-de Sitter (AdS) fifth dimension $z$ to 
the hadron light-front wave functions (LFWFs) in physical space-time. 
This approach has been successfully applied to the description of the 
mass spectrum of mesons and baryons (e.g. reproducing the Regge trajectories), 
the pion leptonic constant, the electromagnetic form factors of pion 
and nucleons, etc.~\cite{Brodsky:2007hb}-\cite{Gutsche:2011vb}. 
The mapping that allows to relate AdS modes to a LFWF is obtained by 
matching certain matrix elements (e.g. the electromagnetic pion form factor) 
in the two approaches - string theory in AdS and Light-Front QCD in 
Minkowski space-time.

The same idea can be used to calculate the generalized parton 
distributions (GPDs) of the nucleon~\cite{Vega:2010ns,Vega:2011ck,Vega:2011fz}, 
which encode important information about the hadronic structure. 
This is a new subject in the application of the gauge/gravity correspondence 
to hadronic properties in the strong coupling limit where QCD cannot 
be used in a straightforward way.

GPDs are of nonperturbative nature and therefore cannot be 
easily calculated from Quantum Chromodynamics (QCD).  We have 
essentially three ways to access the GPDs (for reviews see 
e.g.~\cite{Goeke:2001tz,Ji:2004gf}): extraction from the experimental 
measurement of hard processes, a direct  calculation in the context 
of lattice QCD and different phenomenological models and methods. 
An example of the latter procedure is based on a parametrization of 
the quark wave functions/GPDs using constraints imposed by 
sum rules~\cite{Ji:1996nm,Radyushkin:1997ki}, which relate the 
parton distributions to nucleon electromagnetic form factors 
(some examples of this procedure can be found e.g. 
in~\cite{Diehl:2004cx,Guidal:2004nd,Selyugin:2009ic}). 

In a previous work~\cite{Vega:2010ns} we showed how to obtain the quark GPDs of the 
nucleon in a soft-wall model. We applied a matching procedure 
similar to the one used in LFH applications. 
In this paper we study helicity-independent GPDs in the hard-wall 
holographical model suggested and developed in 
Refs.~\cite{HW_Refs,Hard_Wall_V}, 
and then applied for the nucleon sector in~\cite{Abidin:2009hr}. 

In general, soft-wall models have some advantages compared to 
hard-wall models. In particular, the hadronic mass spectrum in 
soft-wall models has Regge-like behavior, and most of the calculations can be 
done analytically. But, as can be seen 
e.g. in~\cite{Abidin:2009hr}, in hard-wall models the obtained nucleon
form factors are in better agreement with data. This makes it worthwhile 
to consider the GPDS in the context of holographic hard-wall 
models. Additionally, we would like to discuss the high momentum transfer 
limit ($Q^{2} \to \infty$) for the physical quantities of 
our calculations. Note that in this limit the soft and hard-wall 
approaches have the same behavior: the electromagnetic probe propagating 
in the AdS space decouples from the dilaton. Something similar happens in the 
case of the pion form factor~\cite{Brodsky:2007hb}. 

In this work we consider a procedure similar to the one discussed 
in~\cite{Vega:2010ns} for the soft-wall case, i.e. we perform a matching 
of the nucleon electromagnetic form factors considering two main ideas: 
we use sum rules, derived in QCD~\cite{Ji:1996nm,Radyushkin:1997ki}, which 
contain the GPDs for the valence quarks, and we consider specific integral 
representations  obtained in the AdS/QCD hard-wall 
model~\cite{Abidin:2009hr}. As result of the matching we obtain expressions 
for the nonforward parton densities~\cite{Radyushkin:1998rt} 
$H_{v}^{q}(x,t) = H^q(x,0,t) + H^q(-x,0,t)$ and 
$E_{v}^{q}(x,t) = E^q(x,0,t) + E^q(-x,0,t)$ -- flavor combinations of 
the GPDs (or valence GPDs), using information obtained on the AdS side.

The paper is organized as follows. First, in Sec.~II we review the 
sum rules that relate the GPDs to the nucleon form factors. After that, 
in Sec.~III we summarize the main results of Ref.~\cite{Abidin:2009hr} 
obtained for nucleon form factors in the framework of the hard-wall model. 
In Sec.~IV we outline the matching procedure used in ~\cite{Vega:2010ns} 
now applied to the hard-wall case. In Sec.~V we consider 
the high $Q^2$ limit, noticing that we get the same behavior as in the 
soft-wall case. Finally, we draw our conclusions in Sec.~VI.

\section{GPDs and electromagnetic form factors for the nucleon}

The nucleon electromagnetic form factors $F_1^N$ and $F_2^N$ ($N=p, n$ 
correspond to proton and neutron) are conventionally defined by the 
matrix element of the electromagnetic current as 
\eq 
\label{CorrienteEM}
\langle p' | J^{\mu}(0) | p \rangle = \bar{u}(p') 
[ \gamma^{\mu} F_{1}^N(t) + \frac{i \sigma^{\mu \nu}}{2 m_N}  \, 
q_\nu F_{2}^N(t)] u(p),
\en 
where $q = p' - p$ is the momentum transfer; $m_N$ is the nucleon mass; 
$F_1^N$ and $F_2^N$ are the Dirac and Pauli form factors, which are 
normalized to electric charge $e_N$ and anomalous magnetic moment 
$k_N$ of the corresponding nucleon: $F_1^N(0)=e_N$ and $F_2^N(0)=k_N$.  

The sum rules relating the electromagnetic form factors and the GPDs read 
as~\cite{Ji:1996nm,Radyushkin:1997ki,Radyushkin:1998rt}  
\eq
F_{1}^{p}(t) &=& \int\limits_{0}^{1} dx \, \biggl( \frac{2}{3}H_{v}^{u}(x,t) 
                                  - \frac{1}{3}H_{v}^{d}(x,t)\biggr)\,, 
\label{F1P}\\
F_{1}^{n}(t) &=& \int\limits_{0}^{1} dx \, \biggl( \frac{2}{3}H_{v}^{d}(x,t) 
                                  - \frac{1}{3}H_{v}^{u}(x,t)\biggr)\,, 
\label{F1N}\\                              
F_{2}^{p}(t) &=& \int\limits_{0}^{1} dx \, \biggl( \frac{2}{3}E_{v}^{u} (x,t) 
                                  - \frac{1}{3}E_{v}^{d} (x,t)\biggr)\,, 
\label{F2P}\\
F_{2}^{n}(t) &=& \int\limits_{0}^{1} dx \, \biggl( \frac{2}{3}E_{v}^{d} (x,t) 
                                  - \frac{1}{3}E_{v}^{u} (x,t)\biggr)\,. 
\label{F2N}
\en 
Here we restrict to the contribution of the $u$, $d$ quarks and
respective antiquarks, while the presence of the heavier strange and charm quark 
constituents is not considered. 

\section{Electromagnetic form factors for the nucleon in the 
AdS/QCD hard-wall model}

Here we outline the relevant results for the nucleon form factors using 
a hard wall AdS/QCD model as obtained by Abidin and 
Carlson~\cite{Abidin:2009hr}. In this model a cut off $z_{0}$ is introduced 
in the AdS space, which leads to a breaking of the conformal invariance and
thereby simulates confinement. The AdS metric is specified as 
\begin{equation}
\label{Metrica}
 ds^{2} = g_{MN} dx^M dx^N = 
\frac{1}{z^{2}} (\eta_{\mu \nu} dx^{\mu} dx^{\nu} - dz^{2}),
\end{equation}
where $\mu, \nu = 0, 1, 2, 3$; $\eta_{\mu \nu} = {\rm diag}(1,-1,-1,-1)$ 
is the Minkowski metric tensor and $z$ is the holographical coordinate 
running from zero to $z_{0}$. 

The relevant terms in the AdS/QCD action which generate the 
nucleon form factors are~\cite{Abidin:2009hr}: 
\eq 
S &=& \int d^4x \, dz \, \sqrt{g} \, \Big( \bar\Psi \, e_A^M \, 
\Gamma^A \, V_M \, \Psi \, \nonumber\\&+& \frac{i}{2} \, \eta_{S,V} \, 
\bar\Psi \, e_A^M \, e_B^N \, [\Gamma^A, \Gamma^B] \, F^{(S,V)}_{MN} \, 
\Psi \, \Big) \,, 
\en 
where the basic blocks of the AdS/QCD model are defined as: 
$g = |{\rm det} \, g_{MN}|$; $\Psi$ and $V_M$ are the 5D Dirac 
and vector fields dual to the nucleon and electromagnetic fields, 
respectively; $F_{MN} = \partial_M V_N - \partial_N V_M$; 
$\Gamma^A = (\gamma^\mu, - i \gamma^5)$; $e_A^M = z \delta_A^M$ 
is the inverse vielbein; $\eta_{S,V}$ are the couplings constrained 
by the anomalous magnetic moment of the nucleon (see below). 
Here the indices $S,V$ denote isoscalar and isovector contributions 
to the electromagnetic form factors. 

Finally, results for the nucleon form factors in AdS/QCD are given 
in~\cite{Abidin:2009hr}. For the proton we get: 
\eq 
F_{1}^p(Q^2) &=& C_{1}(Q^2) + \eta_{p} C_{2}(Q^2)\,, \label{F1PH}\\
F_{2}^p(Q^2) &=& \eta_{p} C_{3}(Q^2)\,, \label{F2PH}
\en
and for the neutron
\eq
F_{1}^n(Q^2) &=& \eta_{n} C_{2}(Q^2)\,, \label{F1NH}\\
F_{2}^n(Q^2) &=& \eta_{n} C_{3}(Q^2), \label{F2NH}\,.
\en 
In above expressions we have
$\eta_p = (\eta_S + \eta_V)/2$, $\eta_n = (\eta_S - \eta_V)/2$,  
$Q^{2} = - t$ and the $C_{i}$ are integrals defined by: 
\eq 
C_{1}(Q^2)&=&\int\limits_0^{z_0} dz  
\frac{V(Q,z)}{2 z^{3}} [\psi_{L}^{2}(z) + \psi_{R}^{2}(z)]\,, 
\nonumber \\ 
C_{2}(Q^2)&=&\int\limits_0^{z_0} dz  
\frac{\partial_{z} V(Q,z)}{2 z^{2}} [\psi_{L}^{2}(z) - \psi_{R}^{2}(z)]\,,  
\label{Ci}\\ 
C_{3}(Q^2)&=&\int\limits_0^{z_0} dz  
\frac{2 m_{N} V(Q,z)}{z^{2}} \psi_{L}(z) \psi_{R}(z)\,, 
\nonumber 
\en 
where the coordinate $z$ runs from zero to a maximum value 
$z_{0}$ = (0.245 GeV)$^{-1} \simeq$ 0.81 fm. The latter value was fixed in 
Ref.~\cite{Abidin:2009hr} by the $\rho$--meson 
and nucleon masses. The nucleon magnetic moments 
are expressed in terms of $\eta_p$ and $\eta_n$ as 
$\mu_p = 1 + \eta_p C_3(0)$ and $\mu_p = \eta_n C_3(0)$. 

\begin{figure}[ht]
  \begin{tabular}{c}
    \includegraphics[width=2.5 in]{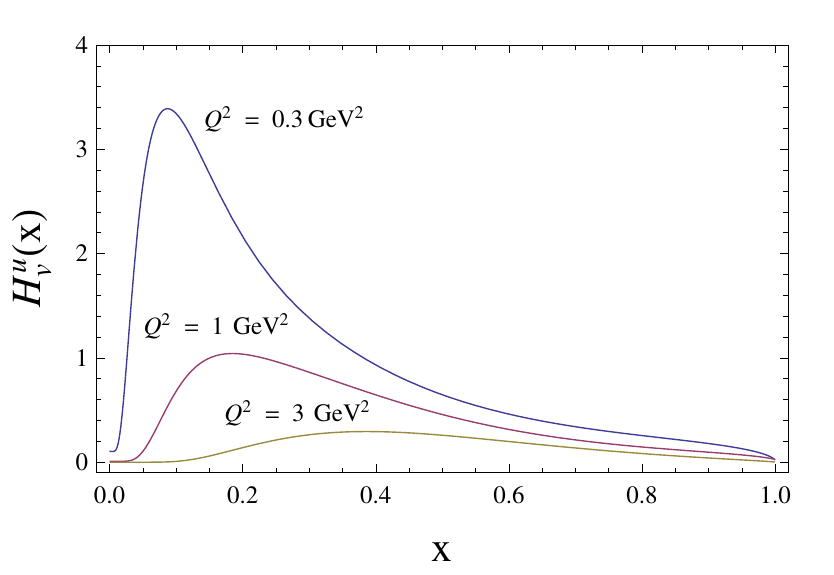}\\
    \includegraphics[width=2.5 in]{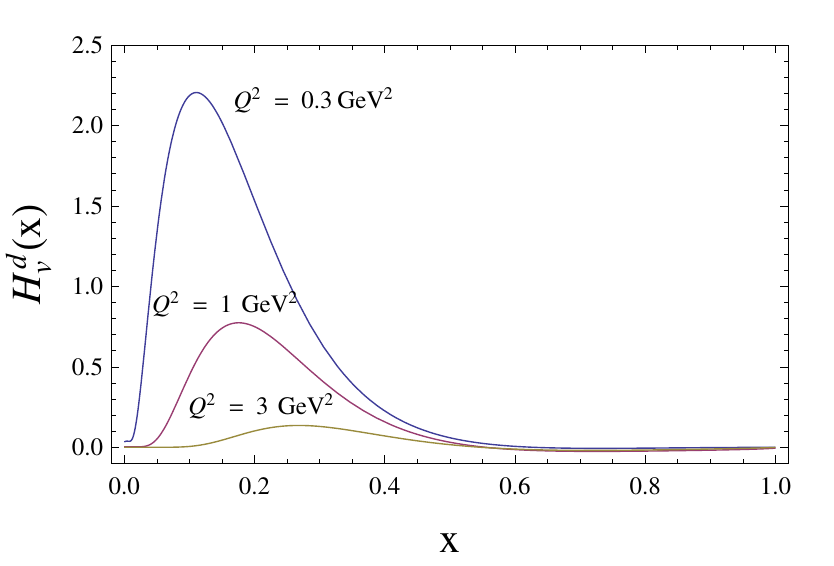}
  \end{tabular}
\caption{$H_{v}^{q} (x)$ in the holographical model for different 
values of $Q^{2}$.}
\end{figure}

\begin{figure}[ht]
  \begin{tabular}{c}
    \includegraphics[width=2.5 in]{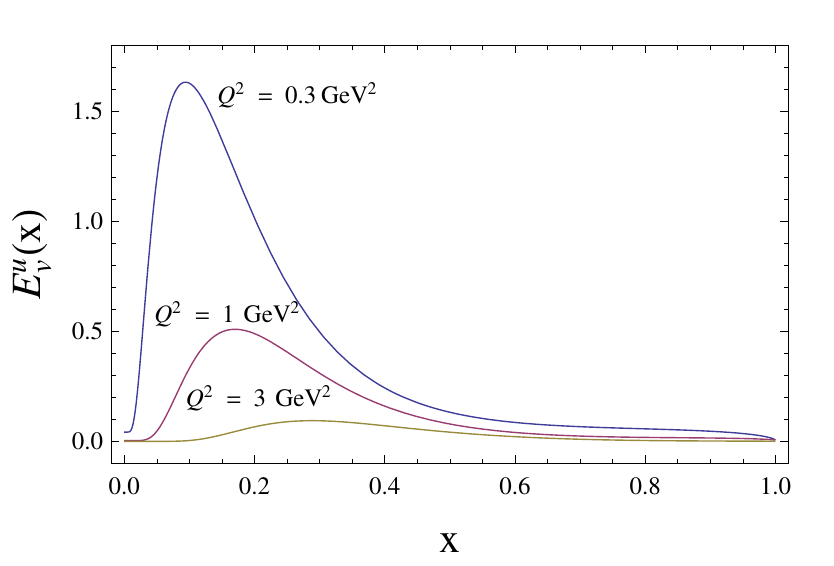}\\
    \includegraphics[width=2.5 in]{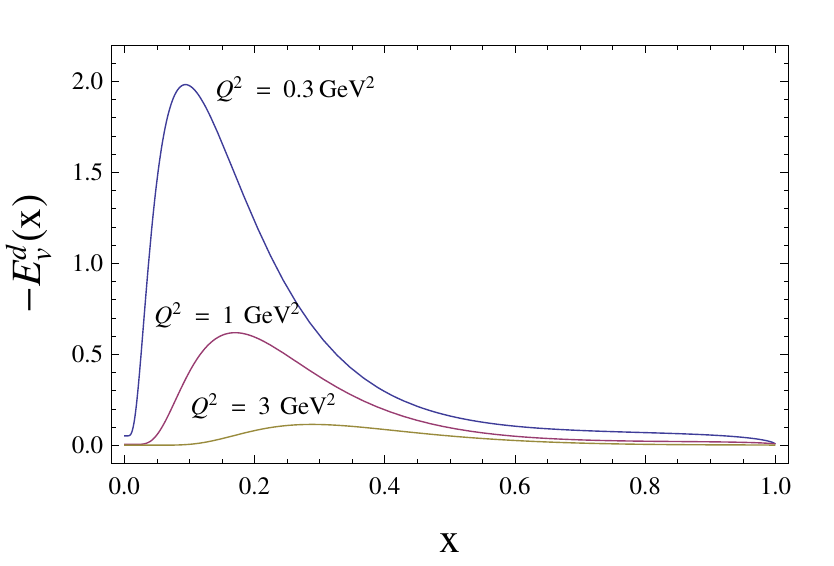}
  \end{tabular}
\caption{$E_{v}^{q} (x)$ in the holographical model for different values 
of $Q^{2}$.}
\end{figure}

According to~\cite{Abidin:2009hr}, $\psi_{L}(z)$ and $\psi_{R}(z)$ are 
\eq 
\psi_{L}(z) &=& \frac{\sqrt{2} z^2 J_2(m_N z)}{z_{0} 
J_2(m_N z_{0})} \label{PsiLRHW1} \\
\psi_{R}(z) &=& \frac{\sqrt{2} z^2 J_1(m_N z)}{z_{0} J_2(m_N z_{0})} 
\label{PsiLRHW2} 
\en 
and for the bulk-to-boundary propagator of the vector field in the 
axial gauge we have~\cite{Hard_Wall_V,Abidin:2009hr}  
\eq
V(Q,z) = Qz \biggl[ \frac{K_{0}(Q z_{0})}{I_{0}(Q z_{0})} I_{1}(Qz) 
        + K_{1}(Q z)  \biggr]\,, \label{V}  
\en 
where $J_\nu$, $I_\nu$, $K_\nu$ are the Bessel and modified Bessel 
functions~\cite{Abr_Stegun}. 
Note that expressions for the nucleon form factors in AdS/QCD can be 
obtained after integration over the variable $z$.
Here we are interested in obtaining GPDs 
using a matching procedure that we will discuss in the next section. 
In this case we follow the formalism developed in Ref.~\cite{Vega:2010ns}. 

Fixing the nucleon magnetic moments $\mu_p = 2.793$ and $\mu_n = - 1.913$
by appropriate choice of the nonminimal couplings $\eta_p=0.448$ and 
$\eta_n = - 0.478$~\cite{Abidin:2009hr} one can get a reasonable description 
of the nucleon form factors and their slopes. In particular, 
the electromagnetic radii of the nucleon are given in the hard-wall model by: 
\eq
\la r^2_E \ra^p &=& - 6 
\biggl(C^\prime_1(0) + \eta_p C^\prime_2(0) - \eta_p \frac{C_3(0)}{4m_N^2} 
\biggr) \,, \nonumber\\ 
\la r^2_E \ra^n &=& - 6 \eta_n 
\biggl( C^\prime_2(0) - \frac{C_3(0)}{4m_N^2}  \biggr) \,, \nonumber\\
\la r^2_M \ra^p &=& 
- 6 \frac{C^\prime_1(0) + \eta_p C^\prime_2(0) + \eta_p C^\prime_3(0)} 
{1 + \eta_p C_3(0)}\,, \nonumber\\
\la r^2_M \ra^n &=& 
- 6 \frac{C^\prime_2(0) + C^\prime_3(0)}{C_3(0)} \,, 
\en
where 
\eq 
C_1^\prime(0) &=& - \frac{1}{8} \int\limits_0^{z_0} \frac{dz}{z} 
\biggl(1 - 2 \log\Big(\frac{z}{z_0}\Big) \biggr) [\psi_{L}^{2}(z) 
+ \psi_{R}^{2}(z)]\,,
\nonumber \\
C_{2}^\prime(0) &=& \frac{1}{2} \int\limits_0^{z_0} \frac{dz}{z^2} 
\log\Big(\frac{z}{z_0}\Big) [\psi_{L}^{2}(z) - \psi_{R}^{2}(z)]\,,
\label{Ci_2}\\
C_{3}^\prime(0) &=& - \frac{m_N}{2} \int\limits_0^{z_0} dz
\biggl(1 - 2 \log\Big(\frac{z}{z_0}\Big) \biggr) \psi_{L}(z) \psi_{R}(z) \,. 
\nonumber
\en 
Note that in the context of AdS/QCD the nucleon charge radii have been discussed 
before in~\cite{Abidin:2009hr}. Here we add the results for the 
magnetic radii.  
Numerical results for the slopes of the nucleon form factors
compare rather well with data:
\eq
\la r^2_E \ra^p &=& 0.829 \ {\rm fm}^2 \ {\rm (our)}\,,
\quad
0.766 \ {\rm fm}^2 \ {\rm (data)}\,, \nonumber\\
\la r^2_E \ra^n &=& - 0.101 \ {\rm fm}^2 \ {\rm (our)}\,,
\quad
- 0.116 \ {\rm fm}^2 \ {\rm (data)}\,, \nonumber\\
\la r^2_M \ra^p &=& 0.756 \ {\rm fm}^2 \ {\rm (our)}\,,
\quad
0.731 \ {\rm fm}^2 \ {\rm (data)}\,, \\
\la r^2_M \ra^n &=& 0.768 \ {\rm fm}^2 \ {\rm (our)}\,,
\quad
0.762 \ {\rm fm}^2 \ {\rm (data)}\,. \nonumber
\en

\section{GPDs and AdS modes in a Soft-Wall model}

As is explained in~\cite{Vega:2010ns}, the matching procedure is based 
on the use of an integral representation for the bulk-to-boundary 
propagator. In the present case, according to (\ref{V}), we need an 
integral representation both for $I_{1} (Q z)$~\cite{Wolfram} and 
$K_{1} (Qz)$ (see Appendix), 
\eq 
\label{I}
I_{1}(Q z) = \frac{2 z Q}{\pi} \int\limits_{0}^{1} dx 
\sqrt{1 - x^{2}} \, {\rm cosh}(x z Q)\,,
\en
\eq 
\label{K}
\hspace*{-.5cm}
K_{1}(Q z) = \frac{z}{Q} \int\limits_{0}^{1} \frac{dx}{(1-x)^{2}} 
\exp(- \frac{Q^2 (1-x)}{4 x} - \frac{z^2 x}{1-x}) \,.
\en 
With these representations it is possible to calculate some of the 
GPDs starting from $V(Q,z)$ and $\psi_{R/L} (z)$. The 
procedure consists in replacing (\ref{V}), using (\ref{I}) and 
(\ref{K}) in (\ref{Ci}), and after performing the $z$ integration, each 
$C_{i}$ expression is now an integral in $x$, which runs from 0~to~1. 
Using this technique in Eqs.~(\ref{F1PH})-(\ref{F2NH}), and 
comparing with Eqs.~(\ref{F1P})-(\ref{F2N}) we can identify the GPDs 
involved in the sum rules. 

The procedure summarized in the last paragraph was applied in the 
soft-wall model to obtain analytical expressions~\cite{Vega:2010ns} 
for some GPDs. Unfortunately here it is not possible to get analytical 
results, although numerical calculations can be done without problems 
using Mathematica. Numerical results for some $Q^{2}$ values are shown 
in Fig.1 and Fig.2. 

\begin{center}
\begin{figure*}[ht]
  \begin{tabular}{c c}
    \includegraphics[width=3.0 in]{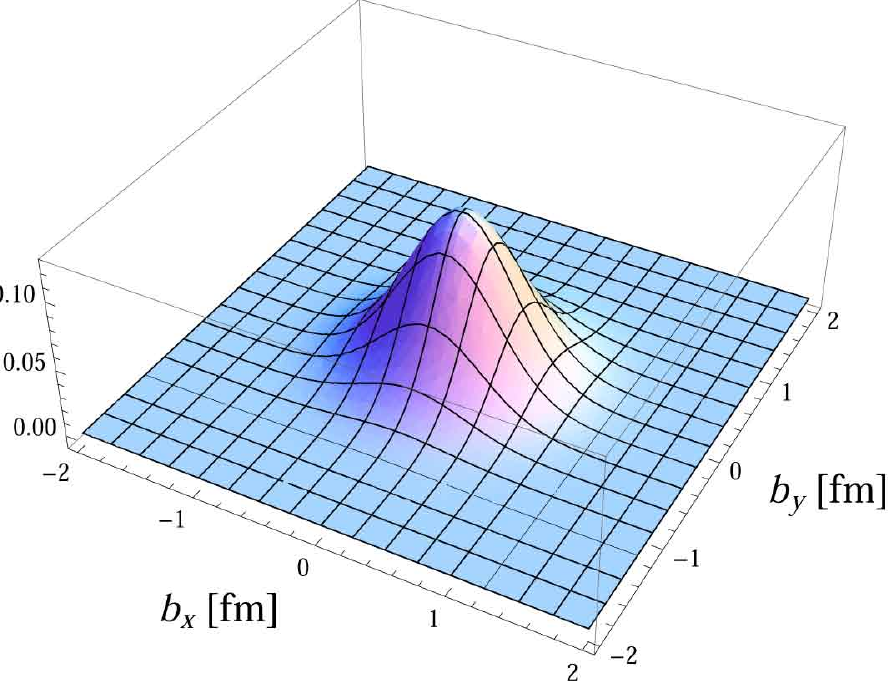}
    \includegraphics[width=3.0 in]{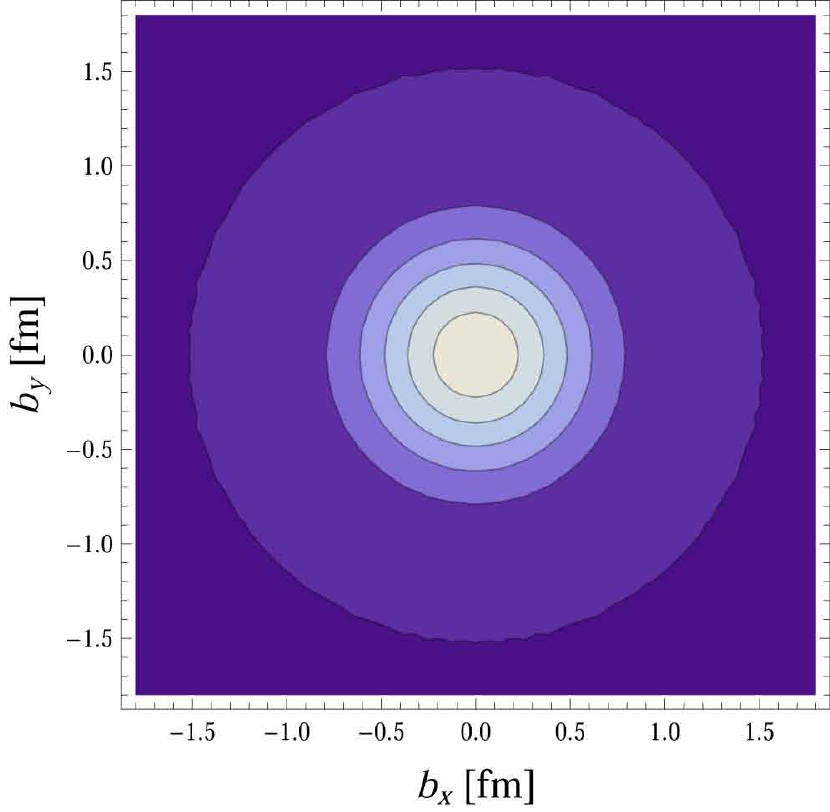} \\
    \includegraphics[width=3.0 in]{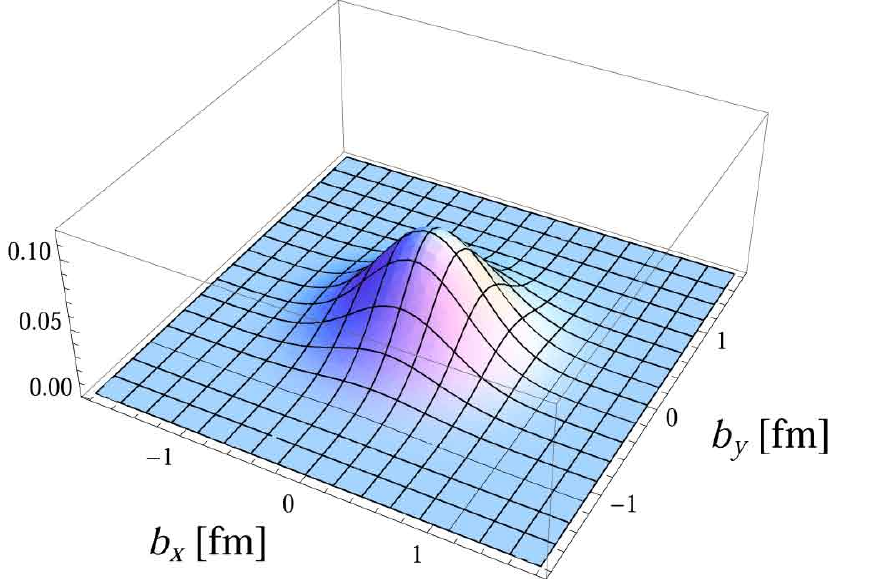}
    \includegraphics[width=3.0 in]{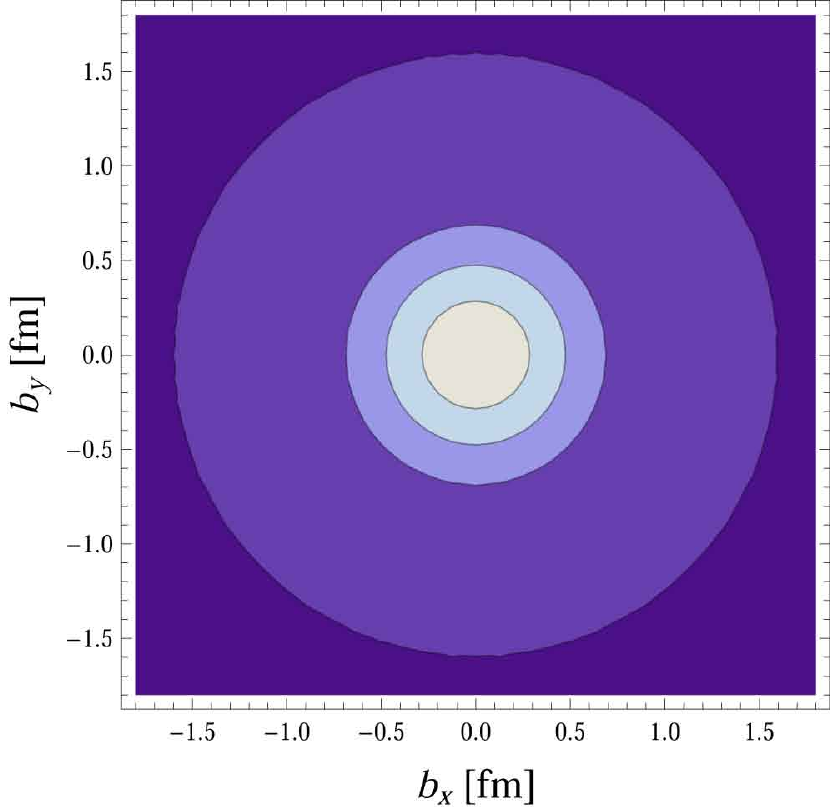}
  \end{tabular}
\caption{Plots for $q(x,\bfb)$. The upper ones correspond to $u(x,\bfb)$ and 
the lower to $d(x,\bfb)$. Both cases are for $x = 0.1$.}
\end{figure*}
\end{center}

\section{Nucleon GPDs in impact space}

Other interesting quantities to consider are the nucleon GPDs 
in impact space. As shown by Burkardt~\cite{Burkardt:2000za, Burkardt:2002hr}, 
the GPDs in momentum space are related to impact parameter 
dependent parton distributions via a Fourier transform. 
GPDs in impact space give access to 
the distribution of partons in the transverse plane, which is 
important for the understanding of the nucleon structure.

\begin{center}
\begin{figure*}[t]
  \begin{tabular}{c c}
    \includegraphics[width=3.0 in]{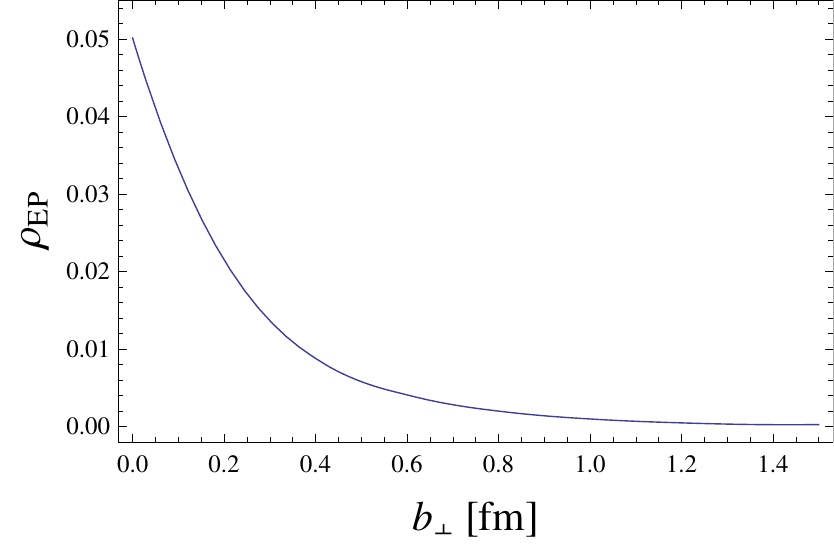}
    \includegraphics[width=3.0 in]{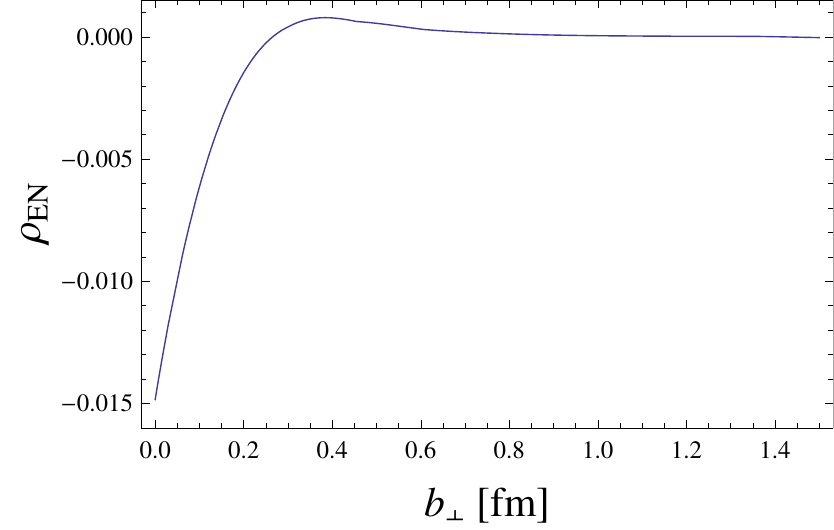} \\
    \includegraphics[width=3.0 in]{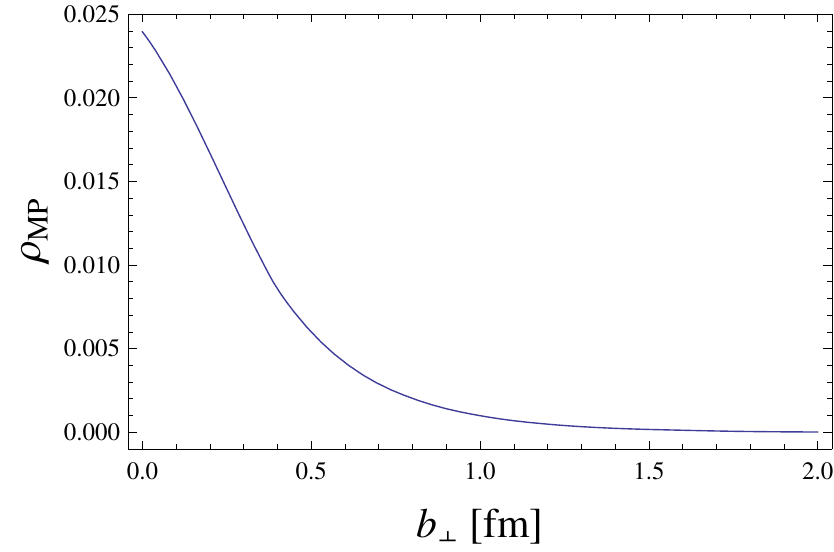}
    \includegraphics[width=3.0 in]{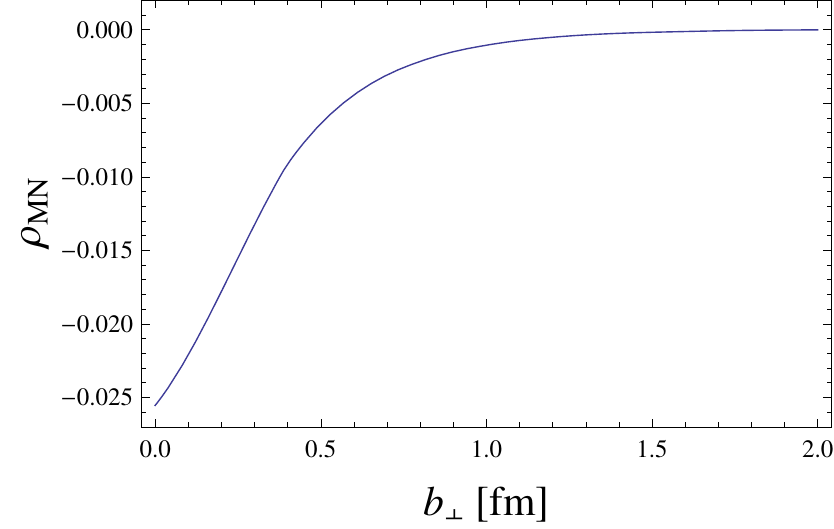}
  \end{tabular}
\caption{Parton charge $(\rho_E(\bfb))$ 
and magnetization $(\rho_M(\bfb))$ densities in  
transverse impact space.}
\end{figure*}
\end{center}

As an example we consider the nucleon GPDs in impact 
space~\cite{Diehl:2004cx,Miller:2007uy,Burkardt:2000za,Burkardt:2002hr} as 
\eq 
q(x,\bfb) &=& \int\frac{d^2\bfk}{(2\pi)^2} H_q(x,\bfk^2) 
e^{-i\bfb\bfk}\,, \nonumber\\
e^q(x,\bfb) &=& \int\frac{d^2\bfk}{(2\pi)^2} E_q(x,\bfk^2) 
e^{-i\bfb\bfk}\,, 
\en 
and the parton charge $(\rho_E(\bfb))$ and magnetization $(\rho_M(\bfb))$ 
densities in transverse impact space 
\eq 
\rho_E(\bfb) &=& \sum\limits_{q} e_q \int\limits_0^1 dx q(x,\bfb) \,, 
\nonumber\\
\rho_M(\bfb) &=& \sum\limits_{q} e_q \int\limits_0^1 dx e^q(x,\bfb) \,. 
\en
Results for both flavors of $q(x,\bfb)$ with x=0.1 are shown in Fig. 3.
In Fig. 4 we give the results for the parton charge and magnetization densities
in transverse impact space.   

\section{Comparison with the soft-wall model in the high $Q^2$ limit}

In the high $Q^2$ limit only the second term in (\ref{V}) is important 
with $V(Q,z) \rightarrow Qz K_{1}(Qz)$. 
Since the Bessel function $K_{1}(Qz)$ decreases 
exponentially in the high $Q^2$ limit,  we only need to consider its
value near $z=0$. 
We therefore need $\psi_{L}(z)$ and $\psi_{R}(z)$ near  $z=0$, i.e.  
\eq 
\psi_{L}(z) &=& \frac{f_N m_N}{4} z^{4}\,, \label{PsiLsw}\\
\psi_{R}(z) &=& f_N z^{3} \,,\label{PsiRsw}
\en 
where
\eq 
\label{fN}
f_N = \frac{m_N}{z_0 J_{2}(m_N z_0) \sqrt{2}}  \,. 
\en 
Note that in the high $Q^2$ limit the integral that appears 
in (\ref{Ci}) has the same form as in the soft-wall case. 
To clarify this point we use some results 
from~\cite{Abidin:2009hr,Vega:2010ns}.

In the soft-wall case the $C_i(Q^2)$ are defined as 
(adding a super-script SW to avoid possible confusion): 
\begin{widetext}
\eq 
C_{1}^{SW}(Q^2)&=&\int\limits_0^\infty dz e^{-\varphi(z)} 
\frac{V(Q,z)}{2 z^{3}} [ (\psi_{L}^{SW}(z))^{2} 
                       + (\psi_{R}^{SW}(z))^{2}]\,,  
\nonumber\\ 
C_{2}^{SW}(Q^2)&=&\int\limits_0^\infty dz e^{-\varphi(z)} 
\frac{\partial_{z} V(Q,z)}{2 z^{2}} [ (\psi_{L}^{SW}(z))^{2} 
                                    - (\psi_{R}^{SW}(z))^{2}]\,,  
\label{CiSW}\\ 
C_{3}^{SW}(Q^2)&=&\int\limits_0^\infty 
dz e^{-\varphi(z)} \frac{2 m_{N} V(Q,z)}{z^{2}} 
\psi_{L}^{SW}(z) \psi_{R}^{SW}(z)\,, \nonumber 
\en 
\end{widetext}
\noindent 
where $\varphi(z)$ is the dilaton field. 
$\psi_{L}^{SW}(z)$ and $\psi_{R}^{SW}(z)$ are the Kaluza-Klein modes 
(normalizable wave functions), which are dual to the left- and right-handed 
nucleon fields:  
\eq \label{PsiLR}
\psi_L^{SW}(z) = \kappa^3 z^4\,, \quad 
\psi_R^{SW}(z) = \kappa^2 z^3 \sqrt{2} 
\en 
and which have the same form as in (\ref{PsiLsw}) and (\ref{PsiRsw}). 
We also need to take the high $Q^2$ limit of the axial gauge vector 
field bulk-to-boundary propagator in the soft-wall model:
\eq
V^{SW} (Q,z) = \Gamma\Big(1 + \frac{Q^{2}}{4 \kappa^{2}}\Big) 
U\Big(\frac{Q^{2}}{4 \kappa^{2}}, 0, \kappa^2 z^2\Big)\,, \label{VSW}
\en  
where $\Gamma(a)$ and $U(a,b,z)$ are the Gamma and Kummer 
functions~\cite{Abr_Stegun}. As can be seen in~\cite{Brodsky:2007hb}, 
we have $V^{SW} (Q,z) \rightarrow zQ K_{1} (z Q)$ for high $Q^2$, 
i.e in this limit 
($Q^{2} \gg 4 \kappa^{2}$) the current decouples from the dilaton field.
Additionally, since we only need the expressions near  $z=0$ we can take 
$e^{- \varphi(z)} = 1$, where we are considering a quadratic dilaton 
$\varphi (z) = \kappa^{2} z^{2}$.  With all these ingredients we can see that 
\eq 
C_{i}^{HW} = C_{i}^{SW}, ~~~~{\rm High}~Q^2~{\rm limit} \,.  
\en 
Note that the parameters of the soft-wall model are the same as those 
used in~\cite{Abidin:2009hr,Vega:2010ns}, i.e. $\kappa = 350$ MeV, 
$\eta_{p} = 0.224$, 
$\eta_{n} = -0.239$, which were fixed in order to reproduce the mass 
$m_N = 2\kappa \sqrt{2}$ and the anomalous magnetic moments of the nucleons.

\section{Conclusions}

We determined numerically the generalized parton distributions of 
nucleons in an AdS/QCD formalism based on a hard-wall scenario 
of conformal symmetry breaking (holographic hard wall model). 
We considered a procedure similar to the one used in some applications 
of LFH and applied previously in~\cite{Vega:2010ns} for the soft-wall 
case. Although soft-wall models in general give a better hadronic Regge 
spectrum than in the hard-wall case,  the latter sometimes is more 
successful in the description of hadronic form factors. It 
provides a strong motivation for the present work. Of course, the 
soft-wall model can also be improved by considering more complicated 
dilaton profiles. 

Another interesting fact is that in the high $Q^2$ limit the form factors 
have the same form in both approaches (hard- and soft-wall models),  
i.e the dilaton decouples in this limit, a property noted 
in~\cite{Brodsky:2007hb} for the pion form factor. So the GPDs must 
be equal in both kinds of holographical approaches at high $Q^2$.

\begin{acknowledgments}

This work was supported by FONDECYT (Chile) under 
Grants No. 3100028 and No. 1100287. This research is also part of 
the Federal Targeted Program ``Scientific and 
sci\-en\-ti\-fic-pe\-da\-go\-gi\-cal personnel of innovative Russia''  
Contract No. 02.740.11.0238. 

\end{acknowledgments}

\appendix

\section{The integral representation used for $K_{1} (zQ)$}

We start by considering the integral (C7) from~\cite{Brodsky:2007hb}
\eq 
K_{\nu} (z) = \frac{z^\nu}{2^{\nu+1}} \, 
\int\limits_{0}^{\infty} \frac{\displaystyle{
e^{-t-\frac{z^2}{4t}}}}{t^{\nu+1}} dt
\en 
that corresponds to an integral representation for the modified Bessel 
function $K_{\nu} (z)$.

Taking $\nu = 1$, replacing $z$ by $zQ$ and considering the following 
change of variable in the integral $t = Q^{2} \frac{1-x}{4x}$, we obtain
\eq 
\hspace*{-.5cm}
K_{1}(Q z) = \frac{z}{Q} \int\limits_{0}^{1} \frac{dx}{(1-x)^{2}} 
\exp( - \frac{Q^2 (1-x)}{4 x} - \frac{z^2x}{1-x} ) 
\en 
which is the integral representation used for $K_{1}(zQ)$.

\end{document}